\documentclass[11pt,a4paper]{article} 
\pdfoutput=1
\usepackage{epsfig}
\usepackage{graphicx}
\usepackage{jheppub}
\usepackage{amsmath}
\usepackage{amsfonts}
\usepackage{amssymb}
\usepackage{graphicx}

\usepackage{color}

\def\beq{\begin{equation}}
\def\eeq{\end{equation}}
\def\beqa{\begin{eqnarray}}
\def\eeqa{\end{eqnarray}}
\newcommand{\nn}{\nonumber}

\def\eq#1{Eq.~(\ref{#1})}
\newcommand\Eqns[2]{Eqs.\,(\ref{#1}) and~(\ref{#2})}

\newcommand{\secn}[1]{section~\ref{#1}}

\newcommand\Appx[1]{appendix~\ref{#1}}

\newcommand{\ord}{{\cal O}}
\newcommand{\RE}{{\rm Re}}

\def\ifm{\ifmmode}

\def\ord{{\cal O} }
\def\cM{{\cal M}}

\definecolor{darkgreen}{rgb}{0.0, 0.4, 0.13}

\newcommand\sss{\scriptscriptstyle}
\newcommand\as{\alpha_{\sss S}} 
\newcommand\gs{g_{\sss S}}
\def\tgs{\tilde\gs}

\def \al #1 {\frac {\as({#1})}{\pi} }
\def \ds #1 {\ooalign{$\hfil/\hfil$\crcr$#1$}}

\def\eps{\varepsilon}
\def\eq#1{Eq.~(\ref{#1})}

\allowdisplaybreaks[1]

\title{Iterating QCD scattering amplitudes in the high-energy limit}

\author[a]{Vittorio Del Duca,\footnote{On leave from INFN, Laboratori Nazionali di Frascati, Italy}}
\affiliation[a]{Institute for Theoretical Physics, ETH Z\"urich, 8093 Z\"urich, Switzerland}

\emailAdd{delducav@itp.phys.ethz.ch}

\abstract{We analyse the high-energy limit of the gluon-gluon
scattering amplitude in QCD, and display an intriguing relation between the finite parts
of the one-loop gluon impact factor and the finite parts of the two-loop Regge trajectory.}


\dedicated{Dedicated to the memory of Lev Nikolaevich Lipatov}

\begin{document}
\begin{flushright}
\vspace*{-25pt}
\end{flushright}
\maketitle
\allowdisplaybreaks 


\section{Introduction}
\label{intro}

The high-energy limit of QCD, in which the squared centre-of-mass energy $s$ is much 
larger than the momentum transfer $|t|$, was pioneered by Lev Lipatov~\cite{Lipatov:1976zz} 
and it is described by the Balitsky-Fadin-Kuraev-Lipatov (BFKL) equation, which resums the large logarithms, $\ln(s/|t|)$,
at leading logarithmic (LL) accuracy~\cite{Fadin:1975cb,Kuraev:1976ge,Kuraev:1977fs,Balitsky:1978ic}, and at
next-to-leading-logarithmic (NLL) accuracy~\cite{Fadin:1998py,Ciafaloni:1998gs,Kotikov:2000pm,Kotikov:2002ab}.
The backbone of the BFKL equation are the quark and gluon scattering amplitudes, which are dominated by the
exchange of a gluon in the $t$ channel, and acquire a ladder structure.
They factorise into quantities, like the impact factors, the Regge trajectory,
and the emission of a gluon along the ladder, a.k.a. the central emission vertex, which constitute
the building blocks of the BFKL equation.

In the last few years, the high-energy limit of QCD and the BFKL equation have undergone an intense scrutiny.
On one hand, it has been realised that the BFKL equation is endowed with a rich mathematical structure.
The functions which describe the analytic structure of the BFKL ladder at LL accuracy in QCD,
and the related ladder in ${\cal N}=4$ super Yang-Mills (SYM) theory,
are single-valued iterated integrals on the moduli space ${\cal M}_{0,4}$ of Riemann spheres 
with four marked points~\cite{Dixon:2012yy,DelDuca:2013lma},
which are single-valued harmonic polylogarithms (SVHPLs)~\cite{BrownSVHPLs}.
In the multi-Regge kinematics, which describe the emission of gluon radiation along the BFKL ladder,
the SVHPLs are generalised to single-valued iterated integrals on the moduli space ${\cal M}_{0,n}$ of Riemann spheres 
with $n$ marked points~\cite{DelDuca:2016lad}, which are single-valued~\cite{BrownSVHPLs,BrownSVMPLs,Brown:2013gia}
multiple polylogarithms (SVMPLs)~\cite{Goncharov:2001iea,Brown:2009qja}.
Further, the functions which describe the analytic structure of the BFKL ladder at NLL accuracy are a generalisation of 
SVHPLs~\cite{DelDuca:2017peo} recently introduced by Schnetz~\cite{Schnetz:2016fhy}.

On the other hand, because scattering amplitudes are infrared divergent, and so are the impact factors, the Regge trajectory,
and the central emission vertex, in which they factorise in the high-energy limit, the study of the scattering amplitudes in
the high-energy limit has benefited from a cross breeding with infrared 
factorisation~\cite{Sotiropoulos:1993rd,Korchemsky:1993hr,Korchemskaya:1994qp,Korchemskaya:1996je,Bret:2011xm,DelDuca:2011ae,Caron-Huot:2013fea,DelDuca:2013ara,DelDuca:2014cya,Caron-Huot:2017fxr,Caron-Huot:2017zfo},
according to which the infrared structure of scattering amplitudes for massless partons is known up to three 
loops~\cite{Aybat:2006wq,Aybat:2006mz,Becher:2009cu,Gardi:2009qi,Almelid:2015jia,Almelid:2017qju}.
The one-loop impact factors~\cite{Fadin:1992zt,Fadin:1993wh,Fadin:1993qb,Lipatov:1996ts,DelDuca:1998kx,Bern:1998sc} and the 
two-loop Regge trajectory~\cite{Fadin:1995xg,Fadin:1996tb,Fadin:1995km,Blumlein:1998ib,DelDuca:2001gu}, which are
building blocks to the BFKL equation at NLL accuracy, have poles in the dimensional regulator $\eps$ in $d=4-2\eps$ dimensions
which can be understood, for the one-loop impact factors, in terms of the one-loop cusp anomalous dimension and the one-loop 
quark and gluon collinear anomalous dimensions~\cite{DelDuca:2013ara,DelDuca:2014cya}, and for the two-loop Regge trajectory 
in terms of the two-loop cusp anomalous dimension~\cite{Korchemskaya:1996je}\footnote{The one-loop central emission 
vertex~\cite{Bern:1998sc,Fadin:1996yv,Fadin:1998sh,DelDuca:1998cx} has not been analysed yet in this fashion.}.

Further, we know that the picture of high-energy factorisation based on one-reggeised-gluon exchange breaks down 
at NNLO accuracy~\cite{DelDuca:2001gu}.
The violation can be explained through infrared factorisation by showing that the real part of the amplitudes 
becomes non-diagonal in the $t$-channel-exchange basis~\cite{Bret:2011xm,DelDuca:2011ae}.
Accordingly, it can be predicted how the violation propagates to higher loops, and the three-loop prediction 
for the violation~\cite{DelDuca:2013ara,DelDuca:2014cya}, which has NNLL accuracy, has been confirmed by the explicit computation of the 
three-loop four-point function of ${\cal N}=4$ SYM~\cite{Henn:2016jdu}. 
In the high-energy factorisation picture, that violation is due to the contribution of the three-Reggeised-gluon 
exchange~\cite{Fadin:2016wso,Caron-Huot:2017fxr}. Thus, although a study of the BFKL ladder 
and firstly of its building blocks at NNLL accuracy is yet to be undertaken, we already have a clear picture of
how the factorisation violations occur at NNLL accuracy.

While we have a precise knowledge of how the infrared poles occur and what they mean in the loop corrections
to impact factors and Regge trajectory, the finite parts of those corrections are treated as free parameters,
which neither infrared nor high-energy factorisation can constrain. At NLL accuracy, they are fixed either by direct 
computation~\cite{Fadin:1992zt,Fadin:1993wh,Lipatov:1996ts,Fadin:1995xg,Fadin:1996tb,Fadin:1995km,Blumlein:1998ib,Fadin:1996yv,Fadin:1998sh} 
or by extracting them from the knowledge of the scattering amplitudes in the exact 
kinematics~\cite{DelDuca:1998kx,Bern:1998sc,DelDuca:2001gu,DelDuca:1998cx}.
The goal of this short note is to alert the reader to the possibility that there might be some organised structure
in the finite parts of impact factors and Regge trajectory as well. We shall discuss amplitudes at NLL accuracy,
so we shall not be concerned with factorisation violations here. 

In \secn{sec:finite}, we focus on  the finite part of the one-loop gluon impact factor, having the finite part of the two-loop Regge trajectory as a target.
In \secn{discuss}, we discuss exponentiation patterns that might help in understanding the findings of \secn{sec:finite}.
In \Appx{AppAnDim}, we recall the coefficients of the cusp and collinear anomalous dimensions, used throughout the note.
In \Appx{AppQIF}, we provide for completeness also the one-loop quark impact factor.

\section{Playing with finite terms}
\label{sec:finite}

In order to illustrate the issue, we shall consider the tree-level amplitude for gluon-gluon scattering
$g_a\, g_b\to g_{a'}\,g_{b'}$, which may be written as \cite{Kuraev:1976ge}, 
\begin{equation}
\cM^{(0) aa'bb'}_{gg\to gg} = 2  s
\left[\gs\, (T^c_G)_{aa'}\, C^{i(0)}(p_a,p_{a'}) 
\right]
{1\over t} \left[\gs\, (T^c_G)_{bb'}\, C^{j(0)}(p_b,p_{b'}) 
\right]\, ,\label{elas}
\end{equation}
where $a, a', b, b'$ represent the colours of the scattering gluons,
and $G$ represents the adjoint representation of $SU(N)$, with
$(T^c_{\sss G})_{ab} = i f^{acb}$, with the colour matrices normalised
in the fundamental representation as
${\rm tr}(T^c_{\sss F} T^d_{\sss F}) = \delta^{cd}/2$.
The coefficient functions $C^{g(0)}$, which yield the leading-order impact factors, 
are given in Ref.~\cite{Kuraev:1976ge} in terms of their spin structure 
and in Ref.~\cite{DelDuca:1995zy} at fixed helicities of the external partons.
Their specific form, though, is immaterial here, since we shall factor the tree amplitude away.

Note that to leading power of $s/t$, amplitudes are dominated by gluon exchange in the $t$ channel,
i.e. only $t$-channel gluon ladders contribute to scattering amplitudes in the high-energy limit.
Note also that the colour component of the amplitude (\ref{elas}) is odd under $s\leftrightarrow u$ exchange,
and it defines the exchange of an antisymmetric octet in the $t$ channel.
However, also the kinematic component of \eq{elas} is odd under $s\leftrightarrow u$ exchange,
due to the overall factor of $s/t$ and since in the high-energy limit $u\simeq -s$.
Thus, the amplitude (\ref{elas}) is even under $s\leftrightarrow u$ exchange, which is a general
feature when only $t$-channel gluon ladders contribute.

In the loop corrections to a scattering amplitude in the high-energy limit, two or more gluons characterise the ladder
exchange in the $t$-channel exchange. Colour-wise, the gluon ladder is then described by the $8 \otimes 8$ colour representation,
which is decomposed as $8 \otimes 8 = \{1 \oplus 8_s \oplus 27\} \oplus [8_a \oplus 10 \oplus \bar{10}]$,
where the term in curly brackets in the direct sum is even under $s\leftrightarrow u$ exchange,
while the term in square brackets is odd. However, to LL accuracy in $\ln(s/|t|)$, only the antisymmetric octet
contributes. In fact, the virtual radiative corrections to eq.~(\ref{elas}) are obtained, to all orders
in $\as$, by replacing \cite{Kuraev:1976ge},
\begin{equation}
{1\over t} \to {1\over t} 
\left({s\over -t}\right)^{\alpha(t)}\, ,\label{sud}
\end{equation}
in eq.~(\ref{elas}), where $\alpha(t)$ is the one-loop Regge trajectory,
which is related to a one-loop transverse-momentum integration.
In dimensional regularization, it can be written as
\begin{equation}
\alpha(t) = \gs^2\, C_A\, {2\over\epsilon} 
\left(\mu^2\over -t\right)^{\epsilon} c_{\Gamma}\, ,\label{alph}
\end{equation}
with $C_A = N_c$, the number of colours, and
\begin{equation}
c_{\Gamma} = {\kappa_\Gamma\over (4\pi)^2}\,, \qquad 
\kappa_\Gamma = (4\pi)^\epsilon\, {\Gamma(1+\epsilon)\,
\Gamma^2(1-\epsilon)\over \Gamma(1-2\epsilon)}\, .\label{cgam}
\end{equation}
The reggeization of the gluon is the fact that higher order corrections to gluon exchange in the $t$ channel can
be accounted for by dressing the gluon propagator with the exponential of \eq{sud}.
Thus, at LL accuracy in $\ln(s/t)$, i.e. considering the corrections of $\ord(\as^n \ln^n(s/t))$,
the amplitude is real, and only the antisymmetric octet contributes to it.

At NLL accuracy in $\ln(s/t)$, i.e. considering the corrections of $\ord(\as^n \ln^{n-1}(s/t))$,
the amplitude develops an imaginary part, however one can show, either within the high-energy limit~\cite{Fadin:2006bj}
or through infrared factorisation~\cite{DelDuca:2011ae}, that only the antisymmetric octet contributes to
the real part of the amplitude. That is the reggeization of the gluon at NLL accuracy.
However, because also virtual corrections to the impact factors contribute to the scattering amplitude
at NLL accuracy, one needs a prescription to disentangle the virtual corrections to the impact factors
in \eq{elas} from the ones that reggeize the gluon (\ref{sud}).
For the antisymmetric octet at NLL accuracy, that prescription is provided by~\cite{Fadin:1993wh}
\beq
\cM^{aa'bb'}_{gg\to gg} = s
\left[\gs\, (T^c_{G})_{aa'}\, C^{g}(p_a,p_{a'}) \right]
{1\over t} \left[\left({-s\over -t}\right)^{\alpha(t)} +
\left({s\over -t}\right)^{\alpha(t)}  \right]
\left[\gs\, (T^c_{\sss G})_{bb'}\, C^{g}(p_b,p_{b'}) \right]\, ,
\label{elasbg}
\eeq
where we analytically continue $\ln(-s) = \ln(s) - i\pi$, for $s > 0$, and
where the impact factors are real and have the loop expansion,
\begin{equation}
C^{g}(t) = C^{g (0)}(t) \left(1 + \sum_{L=1}^\infty \tgs^{2L} C^{g (L)}(\eps) \right)\,
.\label{fullv}
\end{equation}
The Regge trajectory is loop-expanded as,
\begin{equation}
\alpha(t) = \sum_{L=1}^\infty \tgs^{2L} \alpha^{(L)}(\eps)\,
,\label{alphb}
\end{equation}
with the one-loop term given in \eq{alph}. In \Eqns{fullv}{alphb} we have used the rescaled coupling,
\beq
\tgs^2 = \frac{\as}{4\pi} \kappa_\Gamma\, \left({\mu^2\over -t}\right)^{\eps}\, .\label{rescal}
\eeq
Then we write the amplitude (\ref{elasbg}) as a loop expansion proportional to the tree amplitude,
\beq
\cM^{aa'bb'}_{gg\to gg} =  \cM^{(0) aa'bb'}_{gg\to gg}
\left( 1 + \sum_{L=1}^\infty \tgs^{2L} \sum_{i=0}^L M^{(L),i}_{gg\to gg} \ln^i\left(\frac{s}{-t}\right) \right)\, .
\label{elasexpand}
\eeq
The real part of the one-loop term of \Eqns{elasbg}{elasexpand} is
\beq
\RE \left[M^{(1)}_{gg\to gg} \right]
= \alpha^{(1)}(\eps) \ln\left({s\over -t}\right) +\ 2\, C^{g(1)}(\eps)\, .
\label{exp1loop}
\eeq
The Regge trajectory, $\alpha^{(1)}(\eps) = 2C_A/\eps$, 
is in fact independent of the type of
parton undergoing the high-energy scattering process. It is also independent of the infrared
regularisation scheme which is used. Conversely, the one-loop impact factors
are process and infrared-scheme dependent.
They were computed in conventional dimensional 
regularization (CDR)/'t-Hooft-Veltman (HV) schemes in 
Ref.~\cite{Fadin:1992zt,Fadin:1993wh,Fadin:1993qb,Lipatov:1996ts,DelDuca:1998kx,Bern:1998sc}, and
in the dimensional reduction (DR)/ four dimensional helicity (FDH) schemes in Ref.~\cite{Bern:1998sc,DelDuca:1998kx}. To all orders of $\eps$,
the real part of the unrenormalised one-loop amplitude can be written as~\cite{Bern:1998sc},
\beqa
\RE \left[M^{(1)}_{gg\to gg} \right] &=& 
\frac{2C_A}{\eps}\, \left[ - \frac2{\eps} - 2\psi(1-\eps) + \psi(1+\eps) + \psi(1) + \ln\left({s\over -t}\right) \right] \nn\\
&&+ \left( \frac{1-\delta_R\eps}{3-2\eps} - 4\right)\, \frac{C_A}{\eps(1-2\eps)}
+ \frac{2(1-\eps)N_f}{\eps(1-2\eps)(3-2\eps)} \,,
\label{exp1loopalleps}
\eeqa
with $N_f$ the number of light quark flavours, and where, like in \eq{elasexpand}, we have factored out $\tgs^{2}\, \cM^{(0) aa'bb'}_{gg\to gg}$.
In \eq{exp1loopalleps}, we have used the regularisation parameter, $\delta_R =1$ in CDR/HV schemes, $\delta_R =0$ in the DR/FDH  schemes~\cite{Catani:1996pk}. Note also that in the first line of \eq{exp1loopalleps}
the transcendental functions come from one-loop box diagrams, while the rational functions in the second line
come from one-loop bubble diagrams~\cite{Bern:1998sc}.
Using \eq{exp1loop}, we can write the one-loop gluon impact factor to all orders of $\eps$ as,
\beqa
C^{g(1)}(\eps) &=& 
\frac{C_A}{\eps}\, \left[ - \frac2{\eps} - 2\psi(1-\eps) + \psi(1+\eps) + \psi(1) \right] \nn\\
&&+ \left( \frac{1-\delta_R\eps}{3-2\eps} - 4\right)\, \frac{C_A}{2\eps(1-2\eps)}
+ \frac{(1-\eps)N_f}{\eps(1-2\eps)(3-2\eps)} \,.
\label{exp1loopifalleps}
\eeqa
Expanding \eq{exp1loopifalleps} in $\eps$, its $\eps$ poles are accounted for by the cusp anomalous dimension
and by the gluon collinear anomalous dimension~\cite{DelDuca:2014cya}\footnote{Note that in Ref.~\cite{DelDuca:2014cya}
the overall normalisation is different from the one of \eq{rescal}.}, so through $\ord(\eps)$ the one-loop gluon impact factor is
\beqa
C^{g(1)}(\eps) &=& - \frac{\gamma_K^{(1)} }{\epsilon^2}\,C_A + \frac{4\gamma_g^{(1)} }{\epsilon}\,
+ \frac{b_0}{2\eps} + \left( 3\zeta_2 - \frac{67}{18} \right) C_A + \frac{5}9 N_f \nn\\
&& + \left[ \left( \zeta_3 - \frac{202}{27} \right) C_A + {28\over 27} N_f \right] \eps + \ord(\eps^2)\,,
\label{1loopifeps}
\eeqa
with $\gamma_K^{(1)}$ the one-loop coefficient of the cusp anomalous dimension (\ref{eq:k2}), $\gamma_g^{(1)}$ the one-loop coefficient of the 
gluon collinear anomalous dimension (\ref{eq:c1}), and $b_0$ the one-loop coefficient of the beta function (\ref{eq:b0k2}).
Up to the different coefficient of the $\zeta_2$ term, we note that the $\ord(\eps^0)$ term is related to the two-loop cusp anomalous dimension,
$\gamma_K^{(2)}$. Further, we can introduce the quantity, $\mathrm{AD}$\footnote{Up to an overall coefficient, $\mathrm{AD}$ 
coincides with the two-loop anomalous dimension, $G_{eik}$, computed in Ref.~\cite{Erdogan:2011yc}, up to a typo in the sign of the $N_f$ terms
in Eq.~(32).}, defined as,
\beq
\mathrm{AD} = \left( \frac{202}{27} - \zeta_3 \right) C_A - {28\over 27} N_f \,,
\label{newconstant}
\eeq
and we write \eq{1loopifeps} as,
\beq
C^{g(1)}(\eps) = - \frac{\gamma_K^{(1)} }{\epsilon^2}\,C_A - \frac{b_0}{2\eps} + 2\zeta_2 C_A - \gamma_K^{(2)} - \mathrm{AD}\,
\eps + \ord(\eps^2)\,.
\label{1loopifeps2}
\eeq

Since we are interested in exploring the relation of \eq{1loopifeps} to the two-loop Regge trajectory, 
out of the two-loop term in \Eqns{elasbg}{elasexpand}, we only need the real part of the coefficient of the single logarithm,
\beq
\RE \left[M^{(2), 1}_{gg\to gg} \right] = \alpha^{(2)}(\eps) + 2\, C^{g(1)}(\eps)\, \alpha^{(1)}(\eps)\,,
\label{eq:regge2}
\eeq
where $\alpha^{(2)}(\eps)$ is the two-loop Regge trajectory~\cite{Fadin:1995xg,Fadin:1996tb,Fadin:1995km,Blumlein:1998ib,DelDuca:2001gu}.
In CDR/HV, the unrenormalised two-loop trajectory reads,
\beqa
\alpha^{(2)}(\eps) &=& \frac{b_0 }{\eps^2} C_A + \frac{2\gamma_K^{(2)}}{\eps} C_A 
+ \left({404\over 27} - 2\zeta_3\right) C_A^2 - {56\over 27}\, C_AN_f + \ord(\eps)  \label{eq:2loop} \\
&=& 2\, \alpha^{(1)}(2\eps)\, \left[ \frac{b_0}{2\eps} + \gamma_K^{(2)} + \mathrm{AD}\, \eps \right]
+ \ord(\eps)\,, \nn
\eeqa
where in the second line we have used \eq{newconstant}
and factored out the one-loop trajectory $\alpha^{(1)}(\eps) = 2C_A/\eps$, and rewritten it as $2\, \alpha^{(1)}(2\eps)$
for later convenience.

A few comments are in order, when we compare \Eqns{1loopifeps2}{eq:2loop}: as we said, 
in infrared and in high-energy factorisations, the finite part of the one-loop gluon impact factor and the finite part of
the two-loop Regge trajectory are treated as free parameters, i.e. quantities to be determined by an explicit computation. 
However, besides the fact that the $\ord(\eps^0)$ term of the one-loop gluon impact factor (\ref{1loopifeps2})
is related to the two-loop cusp anomalous dimension, up to
the $\zeta_2$ term with a different coefficient\footnote{In a likely related context, SCET in the high-energy limit, 
the two-loop cusp anomalous dimension occurs as the finite term of a one-loop soft exchange~\cite{Rothstein:2016bsq}.},
we observe the even more surprising fact that the
$\ord(\eps)$ term of one-loop gluon impact factor (\ref{1loopifeps2}) exactly reproduces the $\ord(\eps^0)$ term
of the two-loop trajectory (\ref{eq:2loop}).
In fact, apart for the double pole and the different coefficient of the $\zeta_2$ term, one obtains
the two-loop trajectory (\ref{eq:2loop}) by changing sign to \eq{1loopifeps2} and multiplying it by $2C_A/\eps$, i.e. by the one-loop
trajectory.
Such an iterative structure looks surprising: a parton-species dependent one-loop quantity predicts, albeit not fully,
a process independent two-loop quantity\footnote{A similar structure can be distinguished also in the one-loop 
quark impact factor (\ref{1loopqifeps2}), although polluted by additional contributions.}.

\section{Discussion}
\label{discuss}

We have no explanation for 
the intriguing effect we have observed in \eq{1loopifeps2}, which hints at much more structure in the finite parts of scattering
amplitudes in the high-energy limit than ever thought of so far. Although we ignore the physical mechanism which underpins it,
it is worth making a few general remarks about a possible exponentiated form and iterative structure of the QCD scattering
amplitudes in the high-energy limit, including their finite parts. 
However, since infrared factorisation treats the finite parts of impact factor and Regge trajectory as free parameters, thus offering
no guidance to how they might exponentiate, the remarks below are purely heuristic.

We recall that, on other grounds, in Ref.~\cite{DelDuca:2014cya} it was hinted that the impact factors, including their finite
contributions, might exponentiate. Compounded with the already known exponentiation of the Regge trajectory, that amounts, 
to NLL accuracy, to an exponentiation of the (real part of the) QCD one-loop amplitude, $M^{(1)}_{gg\to gg}$.
However, at two loops, although higher orders terms through $\ord(\eps^2)$ would be
present, the square of the one-loop amplitude would yield merely the $2C^{g(1)}\alpha^{(1)}$ term of \eq{eq:regge2},
which is subtracted when we compute the two-loop trajectory. Thus, one would need additional one-loop terms, 
beyond simply the one-loop amplitude.

We note that there is an exponentiation pattern for which the finite part of the Regge trajectory is not a free parameter.
That occurs in planar ${\cal N}=4$ SYM, where the BDS ansatz~\cite{Anastasiou:2003kj,Bern:2005iz} prescribes that 
the four-point scattering amplitude be written as,
\beq
\cM_4 = \cM^{(0)}_4
\exp\left[ \sum_{l=1}^\infty a^l \left( f^{(l)}(\eps) 
M_n^{(1)}(l\eps) + J^{(l)} + \ord(\eps)\right)\right]\, 
,\label{eq:bds1}
\eeq
where $a$ is the 't-Hooft gauge coupling, and with $f^{(l)}(\eps)$ an $l$-loop dependent second-order polynomial in $\eps$,
\beq
f^{(l)}(\eps) = f^{(l)}_0 + \eps f^{(l)}_1 + \eps^2 f^{(l)}_2\, ,\label{eq:flfunct}
\eeq
with $f^{(1)}(\eps)=1$ and where
$f^{(l)}_0$ is proportional to the $l$-loop cusp anomalous 
dimension, and $f^{(l)}_1$
is proportional to the gluon collinear anomalous dimension. 
In \eq{eq:bds1}, $J^{(l)}$ are constants, with $J^{(1)}=0$,
and $M_n^{(L)}(\eps)$ is the $L$-loop colour-stripped
amplitude rescaled by the tree amplitude.

Accordingly, in the high-energy limit of the planar ${\cal N}=4$ SYM 
amplitudes~\cite{Drummond:2007aua,Naculich:2007ub,DelDuca:2008pj,DelDuca:2008jg},
the $l$-loop Regge trajectory can be written as~\cite{DelDuca:2008pj},
\beq
\alpha^{(l)}_{N=4}(\eps) = 2^{l-1}\, \alpha^{(1)}(l\eps)\, \left( f^{(l)}_0 + \eps f^{(l)}_1 \right) + \ord(\eps)\, ,
\label{eq:alphagen}
\eeq
with $\alpha^{(1)}(\eps) = 2/\eps$ the colour-stripped one-loop trajectory of \eq{exp1loop}.
We note a couple of features, which are true to all loops in planar ${\cal N}=4$ SYM:
firstly, \eq{eq:alphagen} is accurate through $\ord(\eps^0)$, thus the second-order term of \eq{eq:flfunct} is irrelevant to it;
secondly, the cusp anomalous dimension fixes the single pole\footnote{Since in ${\cal N}=4$ SYM the beta function is vanishing,
higher order $\eps$ poles do not occur.} of the Regge trajectory, while
the gluon collinear anomalous dimension fixes its finite term.

For $l=2$,  $f^{(2)}_0 = - \zeta_2$ and $f^{(2)}_1 = - \zeta_3$.
The Regge trajectory $\alpha^{(2)}_{N=4}$ (\ref{eq:alphagen}) agrees with the explicit computation of the 
two-loop Regge trajectory in planar ${\cal N}=4$ SYM~\cite{DelDuca:2008pj} and in full ${\cal N}=4$ SYM~\cite{Kotikov:2002ab},
and is in agreement, to $\ord(\eps^0)$, with the terms of highest transcendentality of the QCD two-loop
trajectory (\ref{eq:2loop}).
For $l=3$, $f^{(3)}_0 = 11\zeta_4/2$ and  $f^{(3)}_1 = 6\zeta_5 + 5\zeta_2\zeta_3$.
The Regge trajectory $\alpha^{(3)}_{N=4}$ agrees with the explicit computation of the 
three-loop Regge trajectory in planar ${\cal N}=4$ SYM~\cite{DelDuca:2008pj} and in 
full ${\cal N}=4$ SYM~\cite{Henn:2016jdu}\footnote{If we assign the $N_c$-subleading terms of the coefficient 
of the single logarithm to the factorisation
violations, both for the infrared poles~\cite{DelDuca:2014cya} and for the finite parts. This prescription is
consistent with the fact that the $N_c$-leading term of the coefficient of the single logarithm of full ${\cal N}=4$ SYM
agree with the coefficient of the single logarithm of planar ${\cal N}=4$ SYM.}.

For the QCD amplitudes in the high-energy limit, it is tempting to conjecture an exponentiation pattern like in \eq{eq:bds1},
even more so given that in \eq{eq:bds1} the dependence of $M_n^{(1)}$ on $2\eps$ at two loops stems from the general infrared structure of the 
QCD two-loop amplitudes~\cite{Catani:1998bh,Sterman:2002qn}. 
However, let us also note the differences of the QCD amplitudes in the high-energy limit with respect to planar ${\cal N}=4$ SYM and/or to the BDS ansatz.
Firstly, the finite part of the Regge trajectory (\ref{eq:2loop}) agrees with the QCD gluon collinear anomalous dimension only at the level
of the terms of highest transcendentality. Secondly, the fact that the $\ord(\eps)$ term of the impact factor (\ref{1loopifeps2})
agrees fully with the $\ord(\eps^0)$ term of the Regge trajectory (\ref{eq:2loop}), although present also in planar ${\cal N}=4$ SYM, 
is unaccounted for by the BDS ansatz, which is accurate through $\ord(\eps^0)$.

Finally, if a similar pattern holds at three-loop level, it certainly does in a more intricate way, since the finite part of the three-loop 
Regge trajectory in planar ${\cal N}=4$ SYM~\cite{DelDuca:2008pj} does not occur as such in the higher order terms in $\eps$ 
of the lower-loop impact factor and Regge trajectory. 

That concludes our heuristic exploration of possible exponentiation patterns. As regards the underpinning physical mechanism,
it is conceivable that since the high-energy limit is characterised by two large and disparate hard scales $s \gg -t \gg \Lambda^2$,
the infrared factorisation formulae need be revisited to account for it, with the possible outcome that also some finite
parts of the amplitudes exponentiate, beyond what was envisaged in Ref.~\cite{DelDuca:2014cya}.
Exploring it is left to future analyses.

\section*{Acknowledgements}

The author would like to thank Lance Dixon, Claude Duhr, Giulio Falcioni, Einan Gardi, Lorenzo Magnea, Bernhard Mistlberger
for stimulating discussions and for a critical reading of the manuscript.

\appendix

\section{Anomalous dimensions}
\label{AppAnDim}

The perturbative expansion of the
cusp anomalous dimension~\cite{Korchemsky:1985xj,Moch:2004pa}, divided by the relevant quadratic Casimir factor $C_i$, is
\beq
\label{hatgammaK}
  \gamma_K (\as)  = \sum_{L=1}^\infty \gamma_K^{(L)} \left( \frac{\alpha_s}{\pi} \right)^L\,,
\eeq
with coefficients
\beq
\gamma_K^{(1)} = 2\,,\qquad \gamma_K^{(2)} = \left( \frac{67}{18} - \zeta_2 \right) C_A - \frac{5}{9} N_f\,.
\label{eq:k2}
\eeq

The perturbative expansion of the collinear anomalous dimension is
\beq
\label{collad}
  \gamma_i (\as)  = \sum_{L=1}^\infty \gamma_i^{(L)} \left( \frac{\alpha_s}{\pi} \right)^L\,, \qquad i = q, g\,,
\eeq
with coefficients,
\beq
\gamma_g^{(1)} = - \frac{b_0}{4}\,, \qquad \gamma_q^{(1)} = - \frac{3}4 C_F\,,
\label{eq:c1}
\eeq
where $b_0$ is the coefficient of the beta function,
\beq
b_0 = \frac{11C_A - 2N_F}3\,.
\label{eq:b0k2}
\eeq
As customary in the literature, the expansion in \Eqns{hatgammaK}{collad} is in $\alpha_s/\pi$,
while the impact factor (\ref{fullv}) and the Regge trajectory (\ref{alphb}) are expanded as in \eq{rescal},
however the difference is understood, and should not generate any confusion.

\section{Quark impact factor}
\label{AppQIF}

In CDR/HV, to all orders of $\eps$ the one-loop quark impact factor is~\cite{Fadin:1993qb}
\beqa
C^{q(1)}(\eps) &=& 
\frac{N_c}{\eps}\, \left[ - \frac2{\eps} - 2\psi(1-\eps) + \psi(1+\eps) + \psi(1)  
+ \frac1{1-2\eps} \left( \frac1{4(3-2\eps)} + \frac1{\eps} - \frac7{4} \right) \right] \nn\\
&+& \frac1{N_c}\, \frac1{\eps(1-2\eps)} \left( \frac1{\eps} - \frac{1-2\eps}2 \right)
- \frac{(1-\eps)N_f}{\eps(1-2\eps)(3-2\eps)}\,.
\eeqa
Expanding \eq{exp1loopifalleps} in $\eps$, its $\eps$ poles are accounted for by the cusp anomalous dimension
and by the quark collinear anomalous dimension, so through $\ord(\eps)$ the one-loop quark impact factor is
\beqa
C^{q(1)}(\eps) &=& - \frac{\gamma_K^{(1)} }{\epsilon^2} C_F + \frac4{\eps} \gamma_q^{(1)} + \frac{b_0}{2\eps} - 8 C_F 
+ \left( 3\zeta_2 + \frac{85}{18} \right) C_A - \frac5{9} N_f \nn\\
&+& \left[ - 16 C_F + \left( \zeta_3 + \frac{256}{27} \right) C_A - {28\over 27} N_f \right] \eps + \ord(\eps^2)\,,
\label{1loopqifeps}
\eeqa
with $\gamma_q^{(1)}$ the one-loop coefficient of the quark collinear anomalous dimension (\ref{eq:c1}).
\eq{1loopqifeps} can be also written as,
\beqa
C^{q(1)}(\eps) &=& - \frac{\gamma_K^{(1)} }{\epsilon^2} C_F + \frac4{\eps} \gamma_q^{(1)} 
+ \left[ \frac{b_0}{2\eps} + \gamma_K^{(2)} + \mathrm{AD}\, \eps \right]
\nn\\ &+& (1+4\zeta_2) C_A - 8 C_F + \left[ 2 (1+ \zeta_3) C_A  - 16 C_F \right] \eps + \ord(\eps^2)\,,
\label{1loopqifeps2}
\eeqa
which shows that the same structure found in \Eqns{1loopifeps2}{eq:2loop} can be distinguished also in the one-loop quark impact factor,
although polluted by additional contributions.



\bibliographystyle{JHEP}


\end{document}